\documentclass[12pt,useAMS,usenatbib]{emulateapj}

\usepackage{natbib}

\usepackage{graphicx}
\usepackage{amsmath}
\usepackage{color}

\newcommand{\Msun}{M_{\odot}}

\newcommand{\kpc}{\rm {kpc}}

\newcommand{\art}{\rm ART^{2}}

\newcommand{\Mvir}{M_{\rm{vir}}}
\newcommand{\Vvir}{V_{\rm{vir}}}
\newcommand{\Rvir}{R_{\rm{vir}}}
\newcommand{\Cvir}{C_{\rm{vir}}}

\newcommand{\fesc}{f_{\rm esc}}
\newcommand{\La}{L_{\rm Ly\alpha}}
\newcommand{\A}{\rm \AA}
\newcommand{\Msunyr}{{M_{\odot}~\rm yr^{-1}}}
\newcommand{\ergs}{{\rm erg~s^{-1}}}

\newcommand{\lya}{\ifmmode {\rm Ly}\alpha \else Ly$\alpha$\fi}
\def\msunyr{\ifmmode M_{\odot} ~{\rm yr}^{-1} \else M$_{\odot}$ ~yr$^{-1}$\fi}

\begin{document}                          
%
%

\title{EXTENDED $\lya$ EMISSION FROM INTERACTING GALAXIES AT HIGH REDSHIFTS}    

%
%
\author
{
Hidenobu Yajima\altaffilmark{1,2},
  Yuexing Li\altaffilmark{1,2},
  and Qirong Zhu\altaffilmark{1,2}, 
}

\affil{$^{1}$Department of Astronomy \& Astrophysics, The Pennsylvania State University, 
525 Davey Lab, University Park, PA 16802, USA; yuh19@psu.edu}

\affil{$^{2}$Institute for Gravitation and the Cosmos, The Pennsylvania State University, University Park, PA 16802, USA}


%
%

\begin{abstract}

Recent observations have discovered a population of extended $\lya$ sources, dubbed $\lya$ blobs (LABs), at high redshift $z \sim 2 - 6.6$. These LABs typically have a luminosity of $L \sim 10^{42}-10^{44}~\ergs$, and a size of tens of kiloparsecs, with some giant ones reaching up to $D \sim 100$  kpc. However, the origin of these LABs is not well understood. In this paper, we investigate a merger model for the formation of LABs by studying $\lya$ emission from interacting galaxies at high redshifts by means of a combination of hydrodynamics simulations with three-dimensional radiative transfer calculations. Our galaxy simulations focus on a set of binary major mergers of galaxies with a mass range of $3-7 \times10^{12}~ \Msun$ in the redshift range of $z\sim3 -7$, and we use the newly improved $\art$ code to perform the radiative transfer calculations which couple multi-wavelength continuum, ionization of hydrogen, and $\lya$ line emission. We find that intense star formation and enhanced cooling induced by gravitational interaction produce strong $\lya$ emission from these merging galaxies. The $\lya$ emission appears to be extended due to the extended distribution of sources and gas. During the close encounter of galaxy progenitors when the star formation rate peaks at $\sim 10^3 ~ \Msunyr$, our model produces LABs with luminosity of $L\sim 10^{42}-10^{44}~\ergs$, and size of  $D\sim 10-20 ~\rm kpc$ at $z>6$ and $D\sim 20-50$ kpc at $z \sim 3$, in broad agreement with observations in the same redshift range. Our results suggest that merging galaxies may produce some typical LABs as  observed, but the giant ones may be produced by mergers more massive than those in our model, or a combination of mergers and cold accretion from filaments on a large scale. 

\end{abstract}


%
%

\section{INTRODUCTION}

Star formation in galaxies can produce strong $\lya$ line emission via recombination in H{\sc ii} region in the early phase of the evolution \citep{Partridge67, Charlot93}. Recent narrow-band deep imaging surveys using large-aperture telescopes have detected a large population of $\lya$ emitting galaxies, generally called $\lya$ emitters (LAEs) when the equivalent width (EW) of the $\lya$  line is above $20\;\A$ , at nearly all redshift up to $z=8.6$ \citep[e.g.,][]{Cowie98, Hu98, Malhotra04, Iye06, Gronwall07, Ouchi10, Lehnert10, Kashikawa11, Shibuya11}. It is suggested that most LAEs are compact, young star-forming galaxies (e.g., \citealt{Westra05, Gawiser07, Acquaviva12}, but see \citealt{Kornei10} for a different view).

Meanwhile, a number of extended and bright $\lya$ sources, which are called $\lya$ blobs (LABs), have also been discovered in the redshift range of $z \sim 2 - 6.6$ \citep{Keel99, Steidel00, Steidel11, Reuland03, Palunas04, Matsuda04, Matsuda11, Day05, Saito06, Saito08, Nilsson06, Smith07, Prescott08, Ouchi09a, Yang09, Yang10}. These LABs have a luminosity in the range of $\sim 10^{42} - 10^{44}\, \ergs$, a size from tens of kiloparsecs up to $\sim 100\, \kpc$, much more extended than the LAEs at the same redshift, and some LABs do not seem to have any optical or infrared counterpart.

The origin of the LABs remains a hot debate. One proposal is that they may form from gravitational cooling radiation from accreting gas \citep{Haiman00, Fardal01}, motivated by the observations that some LABs lack visible power source in both optical and infrared \citep{Nilsson06, Smith08}. Recent hydrodynamics simulations show that galaxy evolution is accompanied by accretion of cold gas streams of $T\sim 10^{4}-10^{5}~\rm K$, which penetrate deep inside dark matter halos \citep{Katz03, Keres05, Keres09, Birnboim03, Dekel06, Ocvirk08, Brooks09, Dekel09}. Such  inflow of cold gas may produce strong and extensive $\lya$ emission via collisional excitation, which may result in LABs if there is no dust absorption \citep{Dijkstra09, Faucher09, Goerdt10, Yajima12c}.

However, other observations show that some LABs are associated with regular Lyman break galaxies \citep{Matsuda04}, sub-millimeter and infrared sources which imply active star formation of SFR $\sim 10^{2}-10^{3} M_{\odot} \rm yr^{-1}$ \citep{Chapman01, Geach05, Geach07}, or active galactic nuclei (AGN) and quasars \citep{Bunker03, Weidinger04, Basu04, Geach07, Smith09, Colbert11}. More recently, observations showed asymmetric, filamentary structure in some giant LABs \citep{Matsuda11, Erb11, Rauch11, Rauch12}, which may result from cold accretion or interaction-triggered inflow. 

To explain the extended distribution of $\lya$ emission of LABs, stellar wind from supernovae feedback has been suggested as an effective mechanism \citep{Taniguchi00, Mori06, Dijkstra12}. \citet{Mori06} showed, by means of ultra-high resolution hydrodynamics simulations, that multiple supernovae feedbacks produced extended bubble structure and the compressed gas shell created numerous $\lya$ photons over extended region. However, it  is not clear that triggers the starburst or AGN activity. In addition, most of the theoretical works assume that all $\lya$ photons can escape from the galaxy, but in reality, $\lya$ photons can experience numerous scattering and be absorbed by dust. Moreover, galactic wind is highly suppressed by gravitation potential at $M_{\rm halo} \gtrsim 10^{12}~\Msun$ \citep{Springel05e}, which may not be effective in producing LABs.

In this work, we propose a formation model of LABs from major mergers of gas-rich galaxies. In such a merging process, strong gravitational torques create highly condensed gas clumps in the nuclei region and tidal tails, and trigger intense, global star formation (e.g., \citealt{Hernquist89b, Barnes92, Sanders96, Springel05e, Hopkins06}). At higher redshift (e.g., $z \gtrsim 2$), the major merger rate is higher in dense regions and the progenitors are more gas rich (e.g., \citealt{Li07, Hopkins10}), which may produce strong, extended $\lya$ emission from both recombination of ionized gas and collisional excitation of neutral hydrogen \citep{Yajima12b, Yajima12c, Yajima12d}. 
In fact, recent observations of LABs showed multiple star-forming clumps or
galaxies in the regions \citep{Colbert06, Prescott12, Uchimoto12}, 
which suggest a merger origin of these LABs. 

To test this model, we investigate the $\lya$ property of interacting galaxies by combining hydrodynamical simulations with three dimensional radiative transfer (RT) calculations. The simulations follow the evolution of galaxy mergers of different mass and redshift, and the RT calculations uses the newly improved code $\art$ by \cite{Yajima12b}. The $\art$ code couples multi-wavelength continuum, $\lya$ line, and ionization of hydrogen, which is critical to study the $\lya$ and multi-band properties of galaxies. 

The paper is organized as follows. We describe the galaxy simulations in \S2, and the method of radiative transfer calculations in \S3.  
In \S4, we present the results of $\lya$ properties,  which include the photon escape fraction, emergent $\lya$ luminosity, the EW, the line profile, 
$\lya$ surface brightness and the size. We discuss in \S5 the effect of wind and AGN on the $\lya$ properties, and summarize in \S6.
 
%
%

\section{Galaxy Model}

\begin{deluxetable*}{cccccccccc}
\tabletypesize{\scriptsize}
\tablecaption{\label{tab:sims}Simulation Runs with Different Parameters and Feedback Models}
\tablehead{
\colhead{Runs} &
\colhead{$M_{\rm tot}~(\Msun)$\tablenotemark{a}} &
\colhead{$z_{\rm init}$\tablenotemark{b}} &
\colhead{$N_{\rm DM}$\tablenotemark{c}} &
\colhead{$N_{\rm b}$\tablenotemark{d}} &
\colhead{BH} &
\colhead{Wind} &
\colhead{$f_{\rm gas}$\tablenotemark{e}} &
}
\startdata
z9A & $3\times10^{12}$ & $9$ & $6\times10^{5}$ & $4\times10^{5} $ & On & On & 0.98  \\
z6A & $5\times 10^{12}$ & $6$ & $6\times10^{5}  $& $4\times10^{5} $& On & On & 0.9  \\
z4.5A & $7\times 10^{12}$ & $4.5$ & $6\times10^{5}$  &$ 4\times10^{5} $& On & On & 0.6  \\
z4.5B & $7\times 10^{12}$ & $4.5$ & $6\times10^{5}  $& $4\times10^{5} $& Off & On & 0.6  \\
z4.5C & $7\times 10^{12}$ & $4.5$ & $6\times10^{5}  $& $4\times10^{5}$ & On & Off & 0.6  
\enddata

\tablenotetext{a}{Total mass of the merging system.}
\tablenotetext{b}{Initial redshift of the simulation.}
\tablenotetext{c}{Total number of dark matter particles.}
\tablenotetext{d}{Total number of star and gas particles.}
\tablenotetext{e}{Gas mass fraction to dark matter normalized by {\it WMAP-7} year data.};
\end{deluxetable*}

In order to determine the role mergers play in producing LABs, we perform a set of hydrodynamics simulations of binary, equal-mass, major mergers of galaxies. The galaxy is constructed based on the model of \cite{Mo98}, which consists of a dark matter halo, a disk of gas and stars, and a seed black hole of $10^5\, \Msun$, using a well-tested method (\citealt{Hernquist93, Springel99, Springel00, Springel05d}). We follow the procedures of \cite{Li07} to generate the galaxy progenitors, the properties of which, including the virial mass $\Mvir$, virial radius $\Rvir$ and halo concentration $\Cvir$, are scaled appropriately with redshift.

\begin{eqnarray}
  \Mvir & = & \frac{\Vvir^{3}}{10GH(z)}\,, \\
  \Rvir & = & \frac{\Vvir}{10H(z)} \,, \\
  H(z)  & = & H_0\left[\Omega_{\Lambda} +(1-\Omega_{\Lambda}-\Omega_{\rm
    m})(1+z)^2+\Omega_{\rm m}(1+z)^3\right]^{1/2}\,, \\
  \Cvir & = & 9 \left[\frac{\Mvir}{M_0}\right]^{-0.13}
  \left(1+z\right)^{-1} \,,
\end{eqnarray}
\noindent
where $G$ is the gravitational constant, and $M_0 \sim 8 \times 10^{12} h^{-1} M_{\sun}$ is the linear collapse mass at the present epoch.  

The density profile of the dark matter halo follows a Hernquist profile \citep{Hernquist90}, scaled to match that found in cosmological simulations
\citep{NFW}, as described in \citet{Springel05e}. 
\begin{equation}
\rho (r) = \frac{M_{\rm vir}}{2 \pi} 
\frac{a}{r (r+a)^{3}},
\end{equation}
where $a$ is a parameter that relates the \citet{Hernquist90} profile parameters to the appropriate Navarro-Frenk-White halo scale length $R_{s}$
and concentration $C_{\rm vir}~(C_{\rm vir} = R_{\rm vir}/R_{\rm s})$,
\begin{equation}
a = R_{\rm s}\sqrt{2[{\rm ln}(1+C_{\rm vir}) - C_{\rm vir}/(1+C_{\rm vir})]}.
\end{equation}

The exponential disk of stars and gas are then constructed as in \cite{Hernquist93} and \cite{Springel05d}. We assume a baryon fraction of $f_{\rm{b}}=0.17$ for these high-redshift galaxies based on the seventh year {\it Wilkinson Microwave Anisotropy Probe data} ({\it WMAP7}; \citealt{Komatsu11}). The gas fraction of each progenitor is extrapolated from the results of semi-analytical models of galaxy formation 
\citep{Somerville01}, with 100\% gas disks at $z \ge 10$, 90\% at $10 > z \gtrsim 6$, and 60\% at $6 > z \gtrsim 4$. 

The simulations include dark matter, gasdynamics, star formation, black hole growth, and feedback processes, and are performed using the parallel, $N$-body/smoothed particle hydrodynamics (SPH) code GADGET-3, which is an improved version of that described in \cite{Springel01, Springel05e}. GADGET implements the entropy-conserving formulation of SPH \citep{Springel02} with adaptive particle smoothing, as in \cite{Hernquist89a}. Radiative cooling and heating processes are calculated assuming collisional ionization equilibrium \citep{Katz96, Dave99}. Star formation is modeled in a multi-phase interstellar medium (ISM), with a rate that follows the Schmidt-Kennicutt Law (\citealt{Schmidt59, Kennicutt98}). Feedback from supernovae is captured through a multi-phase model of the ISM by an effective equation of state for star-forming gas \citep{Springel03a}. The UV background model of \cite{Haardt96} is used. A self-regulated black hole model is used, in which a spherical Bondi-Hoyle-Lyttleton gas accretion \citep{Bondi52, Bondi44, Hoyle41} with an upper limit of Eddington rate is assumed, and the feedback is in the form of thermal energy injected to the gas, as described in \cite{Springel05a}. 

We also include a model of galactic wind driven by stellar feedback, as introduced by \cite{Springel03a}. We adopt a constant wind velocity of $v_{\rm wind} = 484~\rm{km~s^{-1}}$, a mass-loss rate that is twice of the star formation rate, and an energy efficiency of unity such that the wind carries $100\%$ of the supernova energy. The wind direction is anisotropical, preferentially perpendicular to the galactic disk. This wind model causes an outflow of gas, transporting energy, matter and metals out of the galactic disk in proportion to the star formation rate \citep{Springel03a}. 

Table~1 lists the simulations performed in this work and the related physical and numerical parameters. The three fiducial runs, z9A, z6A, and z4.5A are set up to produce galaxy mergers at redshift $z \sim$ 6, 4, and 3, respectively, for direct comparison with observations. The progenitors of these mergers belong to the massive end at the given redshift. All these three simulations include thermal feedback from supernovae and black holes, and galactic wind. In order to investigate the effects of feedback on the $\lya$ properties of these merging systems, we also run two additional simulations, z4.5B and z4.5C, without black hole or wind, respectively. We adopt the cosmological parameters from the {\it WMAP7} results:  $H_0 = 72$\,km\,s$^{-1}$\,Mpc$^{-1}$ ($h=0.72$), $\Omega_M=0.26$, $\Omega_{\Lambda}=0.74$, $\Omega_b=0.044$, $\sigma_8 = 0.80$, and $n_s = 0.96$.

\section{Radiative Transfer}

The RT calculations are performed using the 3D Monte Carlo RT code, All-wavelength Radiative Transfer with Adaptive Refinement Tree ($\art$), as recently developed by \cite{Yajima12b}. $\art$ was improved over the original version of \cite{Li08}, and features three essential modules: continuum emission from X-ray to radio, $\lya$ emission from both recombination and collisional excitation, and ionization of neutral hydrogen. The coupling of these three modules, together with an adaptive refinement grid, enables a self-consistent and accurate calculation of the $\lya$ properties, which depend strongly on the UV continuum, ionization structure, and dust content of the object. Moreover, it efficiently produces multi-wavelength properties, such as the spectral energy distribution and images, for direct comparison with multi-band observations. The detailed implementations of the $\art$ code are described in \cite{Li08} and \cite{Yajima12b}. Here we focus on the $\lya$ calculations and briefly outline the process. 

The $\lya$ emission is generated by two major mechanisms: recombination of ionizing photons and collisional excitation of hydrogen gas. In the recombination process, we consider ionization of neutral hydrogen by ionizing radiation from stars, active galactic nucleus (AGN), and UV background (UVB), as well as by collisions by high-temperature gas. The ionized hydrogen atoms then recombine and create $\lya$ photons via the state transition $\rm 2P \rightarrow 1S$. The $\lya$ emissivity from the recombination is
\begin{equation}
\epsilon^{\rm rec}_{\alpha} = f_{\alpha } \alpha_{\rm B} h \nu_{\rm \alpha} n_{\rm e} n_{\rm HII},
\end{equation}
where $\alpha_{\rm B}$ is the case B recombination coefficient, and $f_{\alpha}$ is the average number of $\lya$ photons produced per case B recombination. Here we use $\alpha_{\rm B}$ derived in \citet{Hui97}. Since the temperature dependence of $f_{\alpha}$ is not strong, $f_{\alpha} = 0.68$ is assumed everywhere \citep{Osterbrock06}. The product $h\nu_{\alpha}$ is the energy of a $\lya$ photon, 10.2~eV.
 
In the process of collisional excitation, high temperature electrons can excite the quantum state of hydrogen gas by the collision. Due to the large Einstein A coefficient, the hydrogen gas can occur de-excitation with the $\lya$ emission. The $\lya$ emissivity by the collisional excitation is estimated by
\begin{equation}
\epsilon^{\rm coll}_{\alpha} = C_{\rm Ly \alpha} n_{\rm e} n_{\rm HI},
\end{equation}
where $C_{\rm Ly \alpha}$ is the collisional excitation coefficient,
$C_{\rm Ly\alpha} = 3.7 \times 10^{-17} {\rm exp}(- h\nu_{\alpha}/kT) T^{-1/2}~\rm erg\; s^{-1}\; cm^{3}$ \citep{Osterbrock06}.

Once the ionization structure have been determined, we estimate the intrinsic $\lya$ emissivity in each cell by the sum of above $\lya$ emissivity, $\epsilon_{\alpha} = \epsilon^{\rm rec}_{\alpha} + \epsilon^{\rm coll}_{\alpha}$. 

In RT calculations, dust extinction from the ISM is included. The dust content is estimated according to the gas content and metallicity in each cell, which are taken from the hydrodynamic simulation. The dust-to-gas ratio of the Milky Way is used where the metallicity is of Solar abundance, and it is linearly interpolated for other metallicity. We use the stellar population synthesis model of GALAXEV \citep{Bruzual03} to produce intrinsic spectral energy distributions (SEDs) of stars for a grid of metallicity and age, and we use a simple, broken power law for the AGN \citep{Li08}. A \citet{Salpeter55} initial mass function is used in our calculations. 

We apply $\art$ to the above hydrodynamic merger simulations. In our post-processing procedure, we first calculate the RT of ionizing photons ($\lambda \le 912 \;\rm \AA$) and estimate the ionization fraction of the ISM. The resulting ionization structure is then used to run the $\lya$ RT to derive the emissivity. Our fiducial run is done with $N_{\rm ph} = 10^{5}$ photon packets for each ionizing and $\lya$ components, which was demonstrated to show good convergence \citep{Yajima12b, Yajima12c}. The highest refinement of the adaptive grid corresponds to a cell size comparable to the spatial resolution of 20 pc in physical coordinate of the hydrodynamic simulation.

%
%

\section{$\lya$ Properties}

\begin{figure*}
\begin{center}
\includegraphics[scale=0.6]{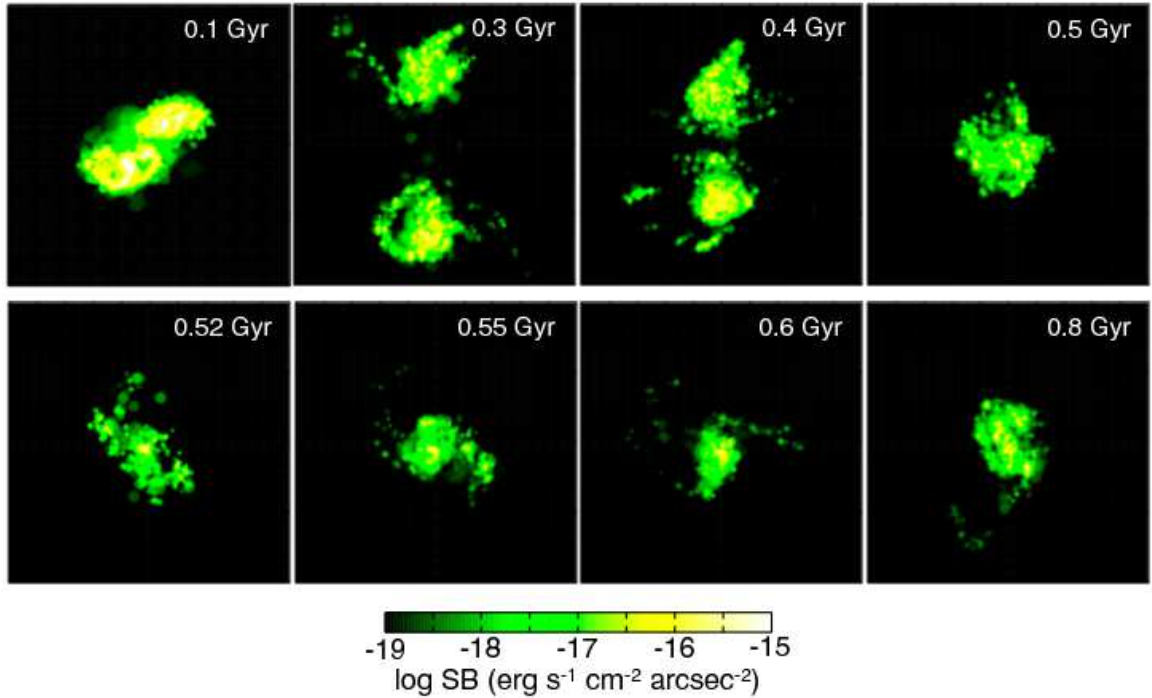}
\caption{
Evolution of the $\lya$ map during the merging event, from the z4.5A simulation. The box size is 100 kpc in physical scale, and the color bar indicates the surface brightness of the $\lya$ flux.
}
\label{fig:img_z4.5}
\end{center}
\end{figure*}

The simulations z9A, z6A, and z4.5A produce mergers at redshift $z \sim$ 6, 4 and 3, respectively. The high-redshift galaxy progenitors in our simulations are gas-rich and compact. The merging systems typically experience a series of separation, close encounter, and final coalescence phases. During the merging event, strong gravitational interactions between the progenitors lead to tidal tails, strong shocks and efficient gas inflow that triggers large-scale starbursts and fuels rapid accretion onto the black holes. Meanwhile, feedback from both supernovae and black holes suppress both star formation and black hole accretion. The resulting $\lya$ emission follows the evolution of the system, as illustrated in Figure~\ref{fig:img_z4.5} from the z4.5A run. 

At $t=0.1$ Gyr after the simulation starts, the galaxy pair is undergoing the first close encounter. The $\lya$ emission extends to a large distance and shows two strong peaks around the two interacting nuclei. As the pair coalesces at $t \sim 0.48 - 0.5$ Gyr, in contrast to a strong spike of starburst, the $\lya$ emission dims and shrinks, and shows holes in the central region of the system, indicating significant absorption. After that, the $\lya$ emission continues to decline and its distribution becomes more compact in the merger remnant. 

\begin{figure}
\begin{center}
\includegraphics[scale=0.45]{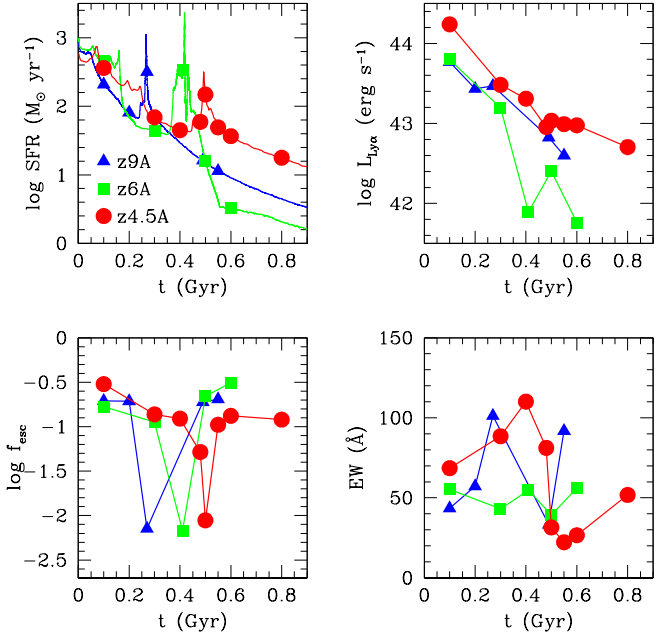}
\caption{Evolution of the physical and $\lya$ properties during the galaxy mergers from the three fiducial simulations, z9A, z6A, and z4.5A, which produce an merging event at $z \sim 6$, 4, and 3, respectively. From top to bottom in clockwise direction are:  the star formation rate (upper left panel), the emergent $\lya$ luminosity (upper right), the equivalent width of $\lya$ line (lower right), and the escape fraction of $\lya$ photons (lower left). 
}
\label{fig:flux_z4.5}
\end{center}
\end{figure}

Figure~\ref{fig:flux_z4.5}  shows the star formation history and the resulting $\lya$ properties of the three fiducial simulations, z9A, z6A, and z4.5A, respectively. Higher redshift system merges faster due to smaller dynamical time scale. All simulations show intense star formation during the merging process,  which in general declines with time except a strong peak at the time of the final coalescence, with rates ranging from a few to $\sim 10^{3} \;\rm \Msunyr$. 

The majority of the $\lya$ photons are absorbed by dust in these merging systems. The escape fractions of $\lya$ photons, $\fesc$, are below 30\%, and during the final coalescence phase, they fall to deep dips of $\fesc \sim 1\%$. This is caused by the sudden increase of absorption from the dust produced by the starburst, as well as a sudden increase of the scattering optical depth in the shocked and compressed gas regions. As a result, the $\lya$ luminosity declines with time, in particular, the z6A run shows a steep drop at the coalescence, even its SFR reaches $\sim 3 \times 10^{3} \;\rm \Msunyr$. The z4.5A run generally has higher luminosity, $\La \sim 10^{42.7}-10^{44.2}~\ergs$, the z9A having $\La \sim 10^{41.7} - 10^{43.8}~\ergs$, and z6A lies in between. The observed luminosity range of LABs in $z \sim 3 -6$ is $\sim 10^{43} - 10^{44}~\ergs$, within the range of our simulations. 

The equivalent width (EW) of $\lya$ line is defined as the ratio of $\lya$ luminosity to the UV flux density at $\gtrsim 1216 \A$, ${\rm EW} = \La / f_{\rm UV}$, where $f_{\rm UV}$ is the mean flux from 1300 $\A$ to 1600 $\A$. As shown in Figure~\ref{fig:flux_z4.5}, the EWs of z4.5A run shows $\sim 25-220 \A$, while z9A has $\sim 40 - 100 \A$, and z6A $\sim 50 \A$. This range is in broad agreement with  observations of EWs of LABs, $\sim 40-200 \A$ \citep[e.g.,][]{Saito06}. 

\begin{figure}
\begin{center}
\includegraphics[scale=0.45]{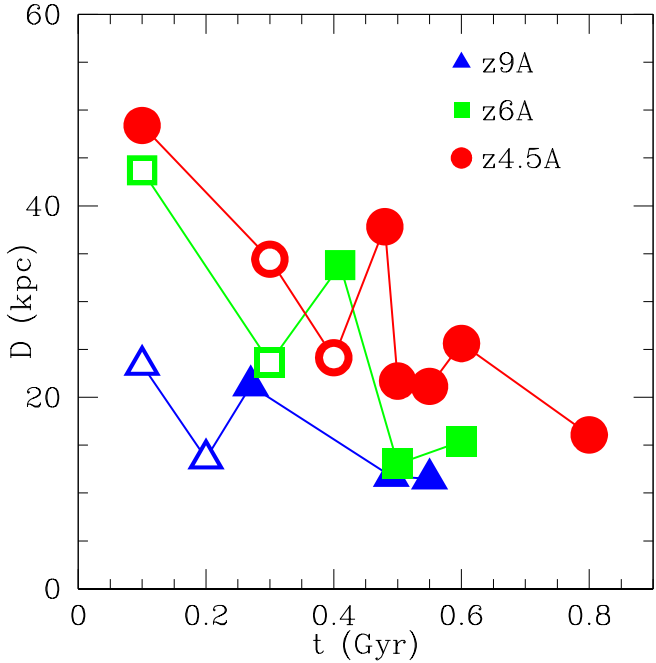}
\caption{
A comparison of the size of LABs above the detection limit from the three fiducial simulations, z9A, z6A, and z4.5A. The flux limit is $2.2\times 10^{-18}~\rm erg\;s^{-1}\;cm^{-2}\;arcsec^{-2}$ from \cite{Matsuda04}. During the merging events, the $\lya$ emission from the two progenitors sometimes overlaps (indicated by filled symbols) and sometimes separates apart (indicated by open symbols). In the separation phase, the size refers to the bigger blob only.  
}
\label{fig:size_z4.5}
\end{center}
\end{figure}

From the three-dimensional $\lya$ map, we measure the size of the blob in a projected plane above a given flux limit. If the two galaxies are separated apart, we measure the size of the bigger one only. Figure~\ref{fig:size_z4.5} gives the size evolution of the $\lya$ blobs from the three fiducial simulations above $2.2\times10^{-18}~\rm erg\; s^{-1}\; cm^{-2}\; arcsec^{-2}$, the detection threshold of \citet{Matsuda04}. The z4.5A run has a size in the range of $D \sim 15 - 50$ kpc during the merging event, z6A in $\sim 15 - 45$ kpc, and z9A in $\sim 10 - 20$ kpc. These results are consistent with observations. For example, the most distant LAB detected at $z=6.6$ by \cite{Ouchi09a} has a luminosity $\La = 3.9\times10^{43}~\ergs$ and a size    $D \sim 17~\rm kpc$, in good agreement with that in the z9A simulation, which represents a merger event in $z \sim 6$. 


%
%
\section{Discussion}

\subsection{Effects of Galactic Wind and AGN Feedback}

It has been suggested that stellar wind from supernovae or AGN feedback may produce the extended $\lya$ distribution of LABs (e.g., \citealt{Taniguchi00, Mori06, Ouchi10}). To examine the effects of feedback on the formation of LABs, we compare simulations with and without feedback. 

In our simulations, we adopt the supernovae feedback model from \cite{Springel03a}, and AGN feedback from \cite{Springel05a}. The feedback from supernovae includes both thermal and kinetic forms. The thermal energy is released to the surrounding ISM, which results in a multi-phase structure. The kinetic feedback is in form of an anisotropic galactic wind perpendicular to the galactic disk. The wind has a constant velocity of $v_{\rm wind} = 484~\rm{km~s^{-1}}$, a mass-loss rate that is twice of the star formation rate, and an energy efficiency of unity such that the wind carries $100\%$ of the supernova energy.  For the AGN feedback, about 5\% of the radiated energy is injected to the gas thermally and isotropically, which would raise the gas temperature and cause an isotropic outflow. 

\begin{figure}
\begin{center}
\includegraphics[scale=0.45]{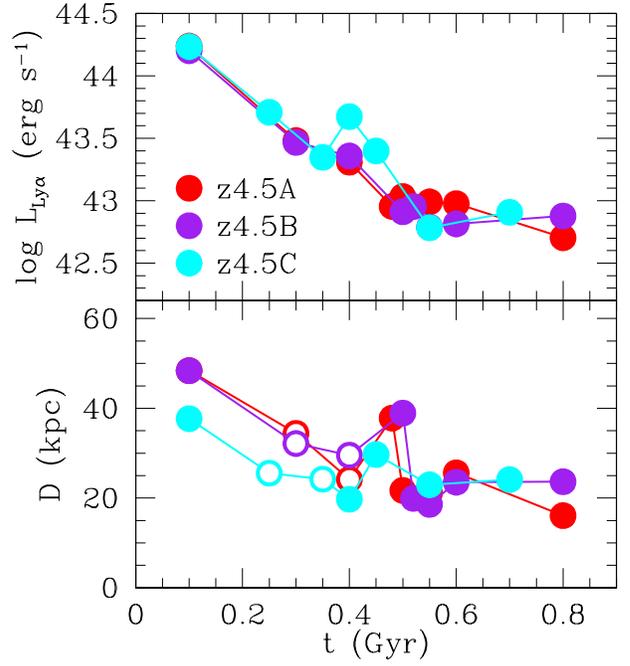}
\caption{
Comparison of  the luminosity ($upper$) and size ($lower$) from simulations 
with different feedback schemes: z4.5A (red, includes both supernovae wind and AGN feedback), z4.5B (purple, no AGN), and z4.5C (cyan, no wind).  Similar to Figure~\ref{fig:size_z4.5}, the flux limit 
for deriving the size
is $2.2\times10^{-18}~\rm erg\; s^{-1}\; cm^{-2}\; arcsec^{-2}$, and the filled symbols indicate that the $\lya$ blobs from progenitors are overlapping, while open symbols indicate that they are separated, and that the size refers to the bigger blob only.
}
\label{fig:size_comp}
\end{center}
\end{figure}

Figure~\ref{fig:size_comp} shows a comparison of the LAB luminosity and size among simulations z4.5A (with both supernovae and AGN feedback), z4.5B (no AGN), and z4.5C (no wind, but with thermal feedback from both supernovae and AGN). While the feedback does not seem to affect significantly the resulting $\lya$ luminosity, the supernovae wind appears to make a remarkable difference in the $\lya$ distribution. Without wind, the $\lya$ is more concentrated, reducing the size by $\sim 10\%$. However, the AGN feedback does not seem have a significant effect on the $\lya$ distribution. 

\subsection{Velocity Width}
Another important feature of LABs is the large velocity width ($\Delta V$) of
the $\lya$ lines. It was suggested that the LABs at $z=3.1$ from
\cite{Matsuda04} have a wide range of velocity width, $\Delta V \sim 500 -
1700~\rm km\;s^{-1}$, while the one detected at $z=6.6$ by \cite{Ouchi09a} has
$\Delta V \sim 250 ~\rm km\;s^{-1}$. 

\begin{figure}
\begin{center}
\includegraphics[scale=0.45]{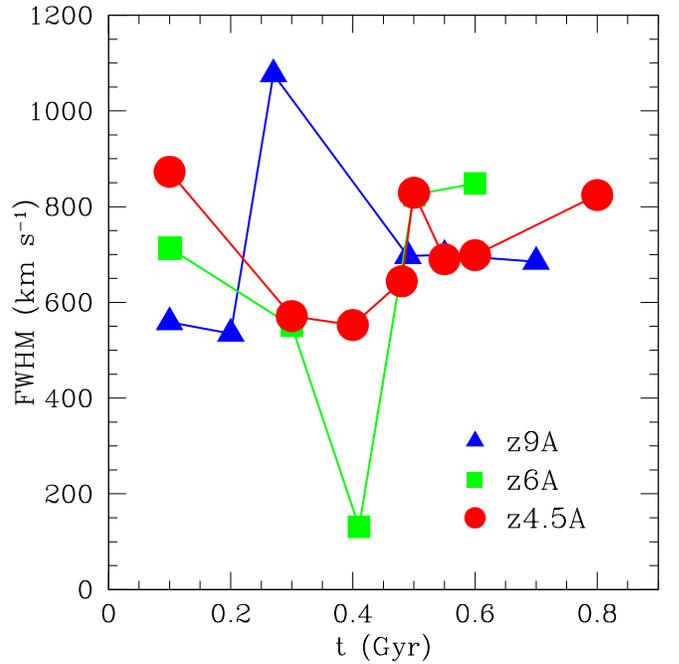}
\caption{
The evolution of velocity width of the $\lya$ line of three different merging
systems in our simulations. The velocity width is measured by the Full Width of Half Maximum
(FWHM) of the emergent $\lya$ line.
}
\label{fig:fwhm}
\end{center}
\end{figure}

Figure~\ref{fig:fwhm} shows the evolution of velocity width $\Delta V$
(measured by the FWHM of the $\lya$ line) from three simulations in our
work. In all cases, $\Delta V$ changes significantly with time as the
interacting galaxies evolve dynamically. During the merging process, large velocity
width is expected when the two progenitors approach to each other, it then
drops after the encounter. After the coalescence, large $\Delta V$ may be maintained by the
rotational motion of gas and the wind feedback due to the active star
formation. In the plot, the peak of z9A run (blue line) corresponds to the
snapshot right before the first encounter, while the dip of the z6A case
(green line) corresponds to a time shortly after the first passage. Overall,
our simulations show a range of $\Delta V \sim 100 - 1100~\rm km\; s^{-1}$ in
these galaxy mergers. This range agrees well with the observations, which
suggests that the large velocity width seen in some LABs may be caused by galaxy
interactions. 

\subsection{Limitations of Our Model}

In this work, the galaxy mergers are idealized and are simulated in
isolation. The mergers constructed at different redshifts show
consistent $\lya$ properties. However, in a cosmological context, galaxies at
different redshift and environment may have different properties such as
geometry, metallicity and dust content, which would significantly affect the
$\lya$ properties. 

In observation, LABs appear more common at high redshifts ($z \sim 3 - 6.6$). 
While some are detected around $z \sim 2$ \citep{Palunas04, Yang09, Yang10,
  Prescott12}, it is rare to find LABs at lower redshift $z < 1$ \citep[e.g.,][]{Keel09}. In our work, however, since we focus on isolated
mergers, we cannot produce a redshift distribution of LABs. Furthermore,   
while our model can produce luminous ($\La \sim 10^{43}-10^{44}~\ergs$) and extended ($D \sim 20 - 50~\rm kpc$) $\lya$ sources at $z \sim 3$
by idealized, binary major mergers, they fall short to reproduce the observed giant LABs with a size of $\sim 100~\rm kpc$ \citep{Matsuda04, Matsuda11}. These giant LABs may form from mergers more massive than the ones we modeled, or from multiple mergers, or from a combination of mergers and accretion of cold gas along the filaments on a large scale. 

In a cold dark matter cosmology, galaxies form hierarchically, with small objects collapsing first and subsequently merging to form progressively
more massive ones. In this picture, massive galaxies form in high overdensity regions where mergers occur frequently \citep{Li07}. The merging event may occur along the filamentary structure of intergalactic medium, and may be accompanied by accretion of cold gas onto the galaxy from the filaments. In these cases, the $\lya$ emission may have very extensive distribution due to scattering along the cold flows and filaments. 

To systematically investigate the formation of LABs, an ideal approach would be to combine multi-wavelength radiative transfer calculations, as those presented here, with full cosmological simulations which follow galaxy formation and evolution at different cosmic epochs. Such large-scale simulations, however, typically lack the high resolutions required to resolve the $\lya$ emission. We will tackle this issue in future work.

%
%

\section{Summary}

By combining hydrodynamical simulations of interacting galaxies and radiative transfer calculations, we investigate the possibility of producing extended $\lya$ blobs from binary, major mergers. We focus on three mergers taking place in the redshift range of $z\sim3 -7$ with a mass range of $3-7 \times10^{12}~ \Msun$. We find that the highly shocked gas regions and intense star formation induced by the gravitational interaction produce strong $\lya$ emission from these merging galaxies. The $\lya$ emission appear to be extended due to the extended distribution of sources and gas.  It has a luminosity of $L\sim 10^{43}-10^{44}~\ergs$  and a size of $D\sim 10-20 ~\rm kpc$ at $z>6$, $D\sim 20-50$ kpc at $z \sim 3$. These results are in good agreement with observations of typical $\lya$ blobs from redshift $z\sim3$ to 6.6. Furthermore, we find that feedback from supernovae or AGN does not have a significant impact on the $\lya$ flux, but the presence of galactic wind from supernovae appear to increase the size of the blobs by $\sim 10\%$. 

We note that our model does not produce giant $\lya$ blobs of $\sim 100$ kpc. This suggests that giant LABs may form from mergers more massive than the ones we presented here, or from multiple mergers, or from a combination of mergers and cold accretion along the filaments on a large scale. To fully account for the $\lya$ sources in these scenarios, one needs to combine high-resolution cosmological hydrodynamics simulations of galaxy formation and evolution, with three-dimensional radiative transfer calculations, which we plan to investigate in a forthcoming paper.
 
 
\acknowledgments

We thank Tom Abel, Mark Dijkstra, Claude-Andr{\'e} Faucher-Gigu{\`e}re, Eric
Gawiser, Caryl Gronwall, Robin Ciardullo and Lars Hernquist for stimulating
discussions and helpful comments. We thank the referee for constructive
suggestions which have helped improve the mannuscript.  
Support from NSF grants AST-0965694 and AST-1009867 is gratefully acknowledged. YL thanks the Institute for Theory and Computation (ITC) at Harvard University where the project was started for warm hospitality. We acknowledge the Research Computing and Cyberinfrastructure unit of Information Technology Services at The Pennsylvania State University for providing computational resources and services that have contributed to the research results reported in this paper (URL: http://rcc.its.psu.edu). The Institute for Gravitation and the Cosmos is supported by the Eberly College of Science and the Office of the Senior Vice President for Research at the Pennsylvania State University.


\end{document}